\documentclass[usenatbib,usegraphicx]{mn2e}

\title[Two-fluid magnetosphere of a pulsar]
{Axisymmetric force-free magnetosphere of a pulsar. II. Transition from the self-consistent two-fluid model}

\author[S. A. Petrova]{S. A. Petrova
\thanks{E-mail: petrova@rian.kharkov.ua}\\
Institute of Radio Astronomy, NAS of Ukraine, Chervonopraporna
Str., 4, Kharkov 61002, Ukraine}

\begin{document}

\date{Received\dots}

\pagerange{\pageref{firstpage}--\pageref{lastpage}} \pubyear{2015}

\maketitle

\protect\label{firstpage}

\begin{abstract}
The self-consistent two-fluid model of the pulsar magnetosphere is considered. We concentrate on the case of vanishingly small inertia of the particles. Our approach allows to obtain the realistic particle distributions sustaining the force-free magnetic field configuration of a monopolar structure. The result differs substantially from the customary picture of the radial speed-of-light motion of massless particles. In our case, the electron and positron constituents follow slightly curved trajectories and are characterized by definite number densities and distinct velocities. The velocity shift is determined by the first-order longitudinal electric field, which appears the necessary ingredient of the self-consistent two-fluid model and implies the parallel conductivity of the order of the inverse particle mass. Our model is believed to be a proper context to describe radiation processes in the pulsar magnetosphere, including both the radio and high-energy emissions. The velocity shift is suggestive of the two-stream instability which may underlie the pulsar radio emission mechanism. The differential rotation of the particle flow may cause the diochotron instability expected to be responsible for the radio subpulse phenomenon. The connection between the radio and high-energy emissions of pulsars is predicted as well.
\end{abstract}


\section{Introduction}

The global structure of pulsar magnetosphere remains the key problem of pulsar theory. A rigid must to interpret a wealth of observational manifestations of pulsars greatly stimulates an interest to the magnetospheric physics. An advanced model of pulsar magnetosphere is thought of as a starting point for understanding the mechanisms of pulsar radio and high-energy emissions, the pulsar wind outflow into the exterior, etc.

In case of pulsars, the presence of copious plasma affects the electromagnetic fields in the magnetosphere, making it necessary to treat the fields and currents self-consistently. Within the framework of the simplest model, the pulsar magnetosphere is presented as an axisymmetric rotator, where the accelerating longitudinal electric field is completely screened by the plasma and the dominant electromagnetic forces are balanced. Such an ideal force-free magnetosphere is described by the well-known pulsar equation \citep{m73,sw73,o74}, which relates the magnetic flux and poloidal current. With the solution in hand, one would automatically obtain the corresponding electric field and charge, reconstructing the general picture of the pulsar magnetosphere. Unfortunately, for the realistic case of a dipolar magnetic field of the neutron star, the proper analytic solution is yet unknown.

The only exact solution of the pulsar equation was guessed for the monopolar magnetic field \citep{m73}. Although the assumption of the stellar magnetic field in the form of a monopole is somewhat unnatural, the versions of the split \citep{m91} and offset \citep{p13} monopole are believed to mimic the salient features of the force-free dipole well beyond the light cylinder. In the ideal force-free case, the monopolar character of the poloidal magnetic field is preserved, whereas the poloidal current gives rise to the azimuthal magnetic field component. As a result, the magnetic field lines are Archimedean spirals, with the projections in the meridional plane being the radial straight lines.

Generally speaking, the force-free consideration involves only the charges and currents rather than the particle distributions sustaining them. For the monopolar case, however, the appropriate particle distribution was guessed as well \citep{m73}. Given that the inertial effects are ignored, the (massless) particles stream radially at a speed of light, following the poloidal magnetic field lines. This result was later on generalized to the case of electromagnetic fields of an arbitrary form \citep[e.g.,][]{gruz08}: The particles also move at a speed of light, and their velocity is directed so as to preclude radiation damping.

Implementation of the numerical methods has led to a marked progress in studying the pulsar magnetosphere. Modeling of the axisymmetric force-free dipole \citep*{ckf99,g05} and taking into account differential rotation of the magnetosphere \citep{c05} were followed by a number of further advances \citep[for a review see, e.g.,][]{s13}. Nowadays, it is a common practice to simulate time-dependent inclined dipole in three dimensions \citep*{s06,kc09,bs10,kck12}, including the effect of finite conductivity of the plasma \citep*{kkhc12,lst12}. In parallel to the force-free description of the pulsar magnetosphere, the MHD approach is also developed \citep*[e.g.,][]{k06,b06,tsl13}. Recently, there was an attempt at including the plasma-producing gaps into the global structure of the pulsar magnetosphere \citep{b14}.

In the numerical simulations of the pulsar magnetosphere, the original force-free result is commonly used as a test for consistency of the numerical procedures. As the current studies are direct or indirect generalizations of the primordial force-free problem, they mostly inherited its difficulties.

Although the particle speed-of-light motion is generally thought of as an inseparable constituent of the force-free picture, it actually appears its weakest point. For the force-free regime to hold the particle energy should be much less than the magnetic field energy. At the same time, the energy of the massless particles streaming at a speed of light remains indeterminate. Furthermore, the force-free region cannot be directly joined to the region of particle acceleration, where the particle mass cannot be neglected and the speed of light cannot be arrived at. And finally, the speed-of-light motion of particles in the force-free case is so degenerate that it precludes any radiative processes, whereas interpretation of the pulsar emission is the main motivation for the magnetospheric studies. The latter difficulty used to be avoided by assuming finite conductivity of the plasma, with the particle mass still being neglected \citep[e.g.,][]{kkhc12,lst12}.

In the present paper, we relax the restriction of massless particles and consider the two-fluid model of the pulsar magnetosphere in order to include the inertial effects. The general picture of the two-fluid model is too complicated and allows only simplistic numerical illustrations \citep{ko09}. We use the analytic approach, address the low-mass limit of the model and obtain the particle distributions sustaining the monopolar force-free configuration. In addition to the customary degenerate speed-of-light solution, there exist realistic electron-positron distributions obeying the two-fluid model at $m\to 0$, where $m$ is the particle mass. Note that actually the small parameter of the problem is of the order of $\Omega/\omega_{G_L}$, where $\Omega$ is the rotation frequency of the pulsar, $\omega_{G_L}\equiv eB_L/(mc)$ the particle gyrofrequency, $B_L$ the magnetic field strength at the light cylinder, $e$ the electron charge and $c$ the speed of light. Thus, our low-mass limit may well be called the strong-field one, as is usually done in the literature \citep[e.g.,][]{gruz08}. Although the latter term is undoubtedly more physically meaningful, we choose the former one in order to stress that our consideration is concerned with the vanishingly small inertial effects.

In the present paper, it is shown that continuity of the plasma flow is satisfied (and, correspondingly, the two-fluid model can be constructed) only if the plasma conductivity $\propto m^{-1}$ is assumed. Then the longitudinal electric field component, which is small as $\bmath{E}\cdot\bmath{B}\propto m$, causes the relative shift in the distributions of electrons and positrons $\propto\bmath{E}\cdot\bmath{B}/m$. Hence, the global force-free picture appears to depend on the details of particle acceleration. This demonstrates the proper way to place the force-free magnetosphere into a more general   context, including the particle high-energy emission. Thus, our paper presents an attempt at consistent extending the force-free model of the pulsar magnetosphere and in this respect continues the ideology of the previous paper of the series \citep{p12}.

Our present consideration yields the simplest non-degenerate description of the pulsar magnetosphere, which may serve as a starting point for further analytic and numerical studies. But even now a number of new important features are unveiled. Firstly, at distances $\sim\gamma_0$ in units of the light cylinder radius (where $\gamma_0$ is the initial Lorentz-factor of the particle motion) the particles are subject to strong acceleration. Thus, the solution of the so-called $\sigma$-problem is already prescribed within the framework of the force-free approximation. Secondly, the shift in the distributions of the particle species possibly implies the two-stream instability, so that the radio emission mechanism may be dictated by the global force-free magnetosphere. Thirdly, the particles perform differential rotation, hinting at the possibility of diochotron instability, which may be responsible for the phenomenon of drifting subpulses. And finally, as the shift in the distributions of the particle species is determined by the accelerating electric field, this may somewhat unexpectedly testify to the physical connection between the radio and high-energy emissions of pulsars.

The plan of the paper is as follows. Section 2 contains the basics of the self-consistent two-fluid model. Its low-mass limit is examined in Sect.~3. In Sect.~4, the implications of our results are discussed. A brief summary is given in Sect.~5.

\section{Basics of the self-consistent two-fluid model} 
\protect\label{s2}

\subsection{General formalism}

Let us consider the stationary axially symmetric magnetosphere of a monopolar structure filled with the copious relativistic electron-positron plasma. We are interested in the self-consistent description of the fields and particle motions in this case. For convenience the electric and magnetic field strengths, $\bmath{E}$ and $\bmath{B}$, are normalized to $M/R_L^3$, where $M\equiv B_\star R_\star^3/2$ is the magnetic moment of the neutron star, $R_L\equiv c/\Omega$ the light cylinder radius, $B_\star$ the magnetic field strength at the neutron star surface, and $R_\star$ the stellar radius. Besides that, the velocities are normalized to $c$, the distances to $R_L$ and the electromagnetic forces to $e$. It is assumed that the electromagnetic force dominates, whereas the particle collisions, pressure, gravitation and radiation damping are negligible. Then the equation of motion for each of the particle species is written as
\begin{equation}
\xi\left (\bmath{v}_\pm\cdot\nabla\right )\gamma_\pm\bmath{v}_\pm=\pm\left (\bmath{E}+\bmath{v}_\pm\times\bmath{B}\right )\,,
\label{eq1}
\end{equation}
where $\bmath{v}_\pm$ are the particle velocities, $\gamma_\pm\equiv(1-v_\pm^2)^{-1/2}$ the corresponding Lorentz-factors, and $\xi$ is the resultant numerical factor, $\xi\propto m$. With the normalizations assumed,
\begin{equation}
\xi=\frac{2mc^4}{eB_\star R_\star^3\Omega^2}=\frac{2\Omega}{\omega_{G_L}}\,,
\label{eq17}
\end{equation}
being twice the ratio of the rotation frequency of the neutron star to the particle gyrofrequency at the light cylinder. The numerical value of $\xi$ will be estimated in the next subsection.

In terms of the particle number densities $n_+$ and $n_-$, the total charge $\rho$ and electric current $\bmath{j}$ (both in units of $e$) read
\begin{equation}
\rho=n_+-n_-\,,
\label{eq2}
\end{equation}
\begin{equation}
\bmath{j}=n_+\bmath{v}_+-n_-\bmath{v}_-\,.
\label{eq3}
\end{equation}
Within the framework of the two-fluid model, the flows of both particle species should be continuous,
\begin{equation}
\nabla\cdot\left (n_\pm\bmath{v}_\pm\right )=0\,.
\label{eq4}
\end{equation}
The electromagnetic fields are given by the Maxwell's equations,
\begin{equation}
\nabla\cdot\bmath{E}=\rho\,,
\label{eq5}
\end{equation}
\begin{equation}
\nabla\times\bmath{B}=\bmath{j}\,,
\label{eq6}
\end{equation}
\begin{equation}
\nabla\times\bmath{E}=0\,,
\label{eq7}
\end{equation}
\begin{equation}
\nabla\cdot\bmath{B}=0\,,
\label{eq8}
\end{equation}
where $\rho$ and $\bmath{j}$ are given by equations~(\ref{eq2})-(\ref{eq3}).

The set of equations (\ref{eq1})-(\ref{eq8}) yields a self-consistent description of the electromagnetic fields and particle motions. The equation of motion (\ref{eq1}) actually presents 6 component equations, equations (\ref{eq2})-(\ref{eq3}), (\ref{eq5})-(\ref{eq8}) contain 8 independent equations, and in equation (\ref{eq4}) only one component is independent, since with equation (\ref{eq6}) the condition $\nabla\cdot\bmath{j}=0$ is fulfilled automatically. In total, there are 15 equations for 14 unknowns, namely, the components of the field strengths $\bmath{E}$ and $\bmath{B}$ and the particle velocities $\bmath{v}_\pm$ along with the particle number densities $n_\pm$. Thus, the set of equations (\ref{eq1})-(\ref{eq8}) is overdetermined. The extra equation can be understood as a restriction on the boundary conditions. In specific cases, the above set of equations may become degenerate.

We start from the simplest case of the force-free monopole. In the spherical coordinate system $(r,\theta,\phi)$ with the axis along the rotational and magnetic axes of a pulsar, the force-free electromagnetic fields read
\begin{equation}
\bmath{B}_0=\left (1/r^2,\,0,\,-\sin\theta/r\right )\,,
\label{eq9}
\end{equation}
\begin{equation}
\bmath{E}_0=\left (0,\,-\sin\theta/r,\,0\right )\,.
\label{eq10}
\end{equation}
Given that the particle inertia is ignored, the equation of motion (\ref{eq1}) is reduced to
\begin{equation}
\bmath{E}+\bmath{v}_\pm\times\bmath{B}=0\,.
\label{eq11}
\end{equation}
Substituting equations (\ref{eq9})-(\ref{eq10}) into equation (\ref{eq11}) yields
\begin{equation}
v_{\theta_\pm}=0,\quad v_{\phi_\pm}=r\sin\theta\left (1-v_{r_\pm}\right )\,.
\label{eq12}
\end{equation}
As is seen from equation (\ref{eq12}), the customary radial speed-of-light motion, $\bmath{v}_\pm=(1,0,0)$, is only a particular case, and the force-free configuration can generally be sustained by the particles with more realistic velocity distributions. Note that at $r\to\infty$ it is necessary that $v_{r_\pm}\to 1$, hinting at breakdown of the force-free regime at large distances.

Using equations (\ref{eq9})-(\ref{eq10}) in equations (\ref{eq5})-(\ref{eq6}) yields the force-free charge and current in the form
\begin{equation}
\rho=j_r=-2\cos\theta/r^2,\quad j_\theta=j_\phi=0\,.
\label{eq13}
\end{equation}
With equations (\ref{eq12})-(\ref{eq13}) equation (\ref{eq3}) becomes degenerate: If its radial component is satisfied, the other two are valid identically. Hence, equations (\ref{eq2})-(\ref{eq3}) are reduced to
\[
n_+-n_-=-2\cos\theta/r^2\,,
\]
\begin{equation}
n_+v_{r_+}-n_-v_{r_-}=-2\cos\theta/r^2\,.
\label{eq14}
\end{equation}
One can see that $v_{r_+}=v_{r_-}=1$ is the trivial solution of the set of equations (\ref{eq14}), which allows any number densities of the particles. The non-trivial solution also exists,
\begin{equation}
n_\pm=\frac{\rho\left (1-v_{r_\mp}\right )}{v_{r_+}-v_{r_-}}\,.
\label{eq15}
\end{equation}
Obviously, $v_{r_\pm}=1$ does not seem a proper choice, since then $\gamma_\pm\to \infty$ and the left-hand side of the equation of motion (\ref{eq1}) contains uncertainty.

The particle distributions obeying equations (\ref{eq12}), (\ref{eq15}) should also satisfy the continuity equation (\ref{eq4}). This leads to the condition
\begin{equation}
\frac{v_{r_-}}{1-v_{r_-}}\frac{1-v_{r_+}}{v_{r_+}}=\mu\,,
\label{eq16}
\end{equation}
where $\mu$ is independent of $r$.

According to the above consideration, if one of the radial velocity components, say $v_{r_+}$, were known, one would immediately obtain the other one from equation (\ref{eq16}) and the corresponding number densities from equation (\ref{eq15}). Generally speaking, the form of $v_{r_+}$ is not arbitrary. We suggest that the physically meaningful form corresponds to the low-mass limit of the complete two-fluid problem including the particle inertia. In Sect.~3, the equation of motion is solved in terms of series in $\xi$ in order to find the proper $v_r$. Before this, we estimate the parameter $\xi$, confirming the applicability of such a procedure in the case considered.

\subsection{Numerical estimates}

We are going to quantify the role of inertial effects in the magnetosphere of a pulsar. Note that in our consideration the particle Lorentz-factor is not normalized and, correspondingly, its characteristic value $\gamma_c$ does not enter $\xi$. Thus, the inertial effects are weak on condition that
\begin{equation}
\xi\gamma_c=\frac{2\Omega}{\tilde{\omega}_{G_L}}\ll 1\,,
\label{eq18}
\end{equation}
where $\tilde{\omega}_{G_L}\equiv eB_L/(\gamma_cmc)$ is the relativistic gyrofrequency at the light cylinder. The numerical estimate of equation (\ref{eq17}) reads
\begin{equation}
\xi=10^{-7}P^2B_{12}^{-1}R_6^{-3}\,,
\label{eq19}
\end{equation}
where $P$ is the pulsar period, $B_{12}\equiv B_\star/10^{12}\mathrm{G}$ and $R_6\equiv R_\star/10^6\mathrm{cm}$. One can see that for typical Lorentz-factors of the secondary plasma, $\gamma_c\sim 10^2-10^3$, the condition (\ref{eq18}) is valid for any conceivable set of pulsar parameters.

The force-free approximation holds as long as the kinetic energy density of the plasma flow is less than the energy density of the magnetic field,
\begin{equation}
\frac{\varepsilon_\mathrm{kin}}{\varepsilon_B}=\frac{2\kappa\left (\rho_\mathrm{GJ}/e\right )\gamma_cmc^2}{B^2/8\pi}\ll 1\,,
\label{eq20}
\end{equation}
where $\kappa$ is the plasma multiplicity factor (i.e. the number of pairs per primary particle) and $\rho_\mathrm{GJ}\equiv B/(2\pi eR_L)$ is the Goldreich-Julian charge density. This can be rewritten in the form
\begin{equation}
\frac{\varepsilon_\mathrm{kin}}{\varepsilon_B}=\frac{8\kappa\gamma_c\Omega}{\omega_{G_L}}\left (\frac{R}{R_L}\right )^3=4\kappa\xi\gamma_c\left (\frac{R}{R_L}\right )^3\ll 1\,,
\label{eq21}
\end{equation}
which allows direct comparison with the condition (\ref{eq18}). Given that $\kappa\gg 1$, at the light cylinder the condition (\ref{eq21}) is stronger than the condition (\ref{eq18}): Even if the force-free approximation is already broken, the inertial effects are still insignificant. For typical pulsar parameters, the region of applicability of the force-free approximation stretches well beyond the light cylinder. Our consideration in the low-mass limit, $\xi\gamma_c\ll 1$, allows to trace the pulsar magnetosphere up to the break of the force-free regime.

\section{The low-mass limit} 
\protect\label{s3}

\subsection{Basic equations}

In the present section, we take into account the particle inertia. According to the results of Sect.~2.2, the inertial effects are believed to be so small that the quantities of the two-fluid model are close to their force-free values with an admissible accuracy. However, within the framework of representation of massless particles, the particle velocities are not determined exhaustively. To obtain the proper force-free velocity distributions one has to consider the first approximation in $\xi$, in which case the solvability condition for the quantities of the first order in $\xi$ yields the zero-order quantities unambiguously. The resultant force-free picture would present the low-mass limit of the self-consistent two-fluid model.

Given that $\xi\ll 1$, the quantities entering the equation of motion (\ref{eq1}) can be presented as
\[
\bmath{v}_\pm=\bmath{v}_{0_\pm}+\xi\bmath{v}_{1_\pm}+\dots\,,
\]
\begin{equation}
\bmath{E}=\bmath{E}_0+\xi\bmath{E}_1+\dots\,,
\label{eq22}
\end{equation}
\[
\bmath{B}=\bmath{B}_0+\xi\bmath{B}_1+\dots\,,
\]
where $\bmath{E}_0$ and $\bmath{B}_0$ are the force-free fields given by equations (\ref{eq9})-(\ref{eq10}) and the components of $\bmath{v}_{0_\pm}$ obey the relations (\ref{eq12}) and the continuity condition (\ref{eq16}); hereafter the subscripts '0' are omitted. Note that, in agreement with the estimates of Sect.~2.2, $\xi\gamma_c\ll 1$ and $\gamma$ is not treated as a large parameter. This is essentially distinct from the traditional approach in the literature, when the analogue of equation (\ref{eq12}) is complemented with the assumption $v^2\equiv 1$ \citep[e.g., equation (1) in][can be obtained in such a way]{gruz08}. 

Making use of equation (\ref{eq22}) in equation (\ref{eq1}) and grouping the terms of order $\xi$ yields
\[
v_{r_\pm}\frac{\partial}{\partial r}\left (\gamma_\pm v_{r_\pm}\right )-\frac{v_{\phi_\pm}^2\gamma_\pm}{r}=\pm\left (E_{1_r}-v_{\phi_\pm}B_{1_\theta}-\frac{\sin\theta}{r}v_{1\theta_\pm}\right ),
\]
\[
\frac{v_{\phi_\pm}^2\gamma_\pm\cos\theta}{r\sin\theta}
\]
\begin{equation}
=\mp\left (E_{1_\theta}-v_{r_\pm}B_{1\phi}+v_{\phi_\pm}B_{1_r}+\frac{\sin\theta}{r}v_{1r_\pm}+\frac{v_{1\phi_\pm}}{r^2}\right )\,,
\label{eq23}
\end{equation}
\[
v_{r_\pm}\frac{\partial}{\partial r}\left (\gamma_\pm v_{\phi_\pm}\right )+\frac{v_{r_\pm}v_{\phi_\pm}\gamma_\pm}{r}=\pm\left (E_{1_\phi}+v_{r_\pm}B_{1_\theta}-\frac{v_{1\theta_\pm}}{r^2}\right ).
\]
Excluding $v_{1\theta_\pm}$ from the first and the third equations of the set (\ref{eq23}) and making use of equation (\ref{eq12}), one can obtain
\[
\frac{\mathrm{d}v_{r_\pm}}{\mathrm{d}x}-\frac{1}{2}\left [\frac{1-x}{v_{r_\pm}}+3x-3(1+x)v_{r_\pm}+(2+x)v_{r_\pm}^2\right ]
\]
\begin{equation}
=\pm\frac{a\sqrt{x}}{2v_{r_\pm}\sin^3\theta}\left [(1-x)+2xv_{r_\pm}-(1+x)v_{r_\pm}^2\right ]^{3/2},
\label{eq24}
\end{equation}
where $x\equiv r^2\sin^2\theta$ and
\[
a\equiv\frac{E_{1_r}}{r^2}-\frac{E_{1_\phi}\sin\theta}{r}-\frac{B_{1_\theta}\sin\theta}{r}\equiv \left (\bmath{E}\cdot\bmath{B}\right )_1
\]
is the first-order longitudinal electric field. Together with the continuity condition (\ref{eq16}), equation (\ref{eq24}) yields the radial velocities of the two particle species and the relation between $\mu$ and $a$.

Given that $a\equiv 0$, the two components of equation (\ref{eq24}) become identical and the solution has the form
\begin{equation}
v_{r_\pm}=w_0\equiv\frac{(1-x)\sqrt{1+C_\pm (x-1)}}{1-x\sqrt{1+C_\pm (x-1)}}\,,
\label{eq25}
\end{equation}
where $C_\pm$ are arbitrary constants with respect to $x$. Note that they still can depend on $\theta$, which enters $x$ as a numerical factor, and in general $C_+\neq C_-$.

As is seen from equation (\ref{eq25}), at $x=0$ we have $w_0=\sqrt{1-C_\pm}$. Then, keeping in mind equation (\ref{eq12}), $C_\pm$ can be recognized as the inverse square of the initial Lorentz-factors of electrons and positrons, $C_\pm\equiv1/\gamma_{i_\pm}^2$. The dependency of $w_0$ on $x$ is presented in Fig.~\ref{f1} by the line without markers. As long as $(x-1)C_\pm\ll 1$, $w_0$ increases slightly, remaining almost unchanged, whereas at $(x-1)C_\pm\sim 1$ it starts growing drastically, so that the applicability condition for the low-mass approximation, $\xi\gamma\ll 1$, is ultimately broken.

Note that, according to equation (\ref{eq14}), the velocities of the two particle species should differ. At the same time, one can see that any choice of $C_+\neq C_-$ in equation (\ref{eq25}) cannot satisfy the continuity condition (\ref{eq16}). Thus, at the assumption $a\equiv 0$, it is impossible to construct the self-consistent two-fluid description of the monopolar case. The presence of the first-order longitudinal electric field appears the necessary constituent of the model. With the Ohm's law $j_\Vert =\sigma E_\Vert$, this implies the plasma conductivity $\sigma\sim\xi^{-1}$. Recall that since $E_\Vert\sim\xi$ and $a\equiv E_\Vert/\xi$, $a$ is generally on the order of $\xi^0$. If one take $a\ll 1$ (with $a\sim\xi^0$), one can expect that the velocities of the particle species differ slightly from each other (cf. equation (\ref{eq24})) and from that given by equation (\ref{eq25}), whereas the continuity condition can hopefully be satisfied. As can be intuitively concluded, it is the case that can be relevant to the pulsar magnetosphere and so it is of a certain interest.

\subsection{The case $\vert v_+-v_-\vert/v_+\ll 1$}

To proceed further it is convenient to introduce the quantities
\begin{equation}
u_\pm\equiv\frac{v_{r_\pm}}{1-v_{r_\pm}}\,,
\label{eq26}
\end{equation}
in which case equation (\ref{eq16}) is simplified to
\begin{equation}
\frac{u_-}{u_+}=\mu\,.
\label{eq27}
\end{equation}
Expressing $u_-$ from equation (\ref{eq27}) and excluding $a$ from the pair of equations (\ref{eq24}), we obtain
\[
\frac{\mathrm{d}u_+}{\mathrm{d}t}\left [\left (\frac{2\mu u_+}{t}-1\right )^{3/2}+\mu^2\left (\frac{2u_+}{t}-1\right )^{3/2}\right ]-\frac{t}{2u_+}\left [\left (\frac{3u_+}{t}-1\right )\right.
\]
\begin{equation}
\left.\times\left (\frac{2\mu u_+}{t}-1\right )^{3/2}+\left (\frac{3\mu u_+}{t}-1\right )\left (\frac{2u_+}{t}-1\right )^{3/2}\right ]=0,
\label{eq28}
\end{equation}
where $t\equiv x-1$. Generally speaking, the substitution $y\equiv u_+/t$ would allow to integrate equation (\ref{eq28}) by separating variables. However, the resultant implicit solution appears too complicated and practically inapplicable. Therefore we find it useful to address the relevant simplification.

\begin{figure*}
\includegraphics[width=190mm]{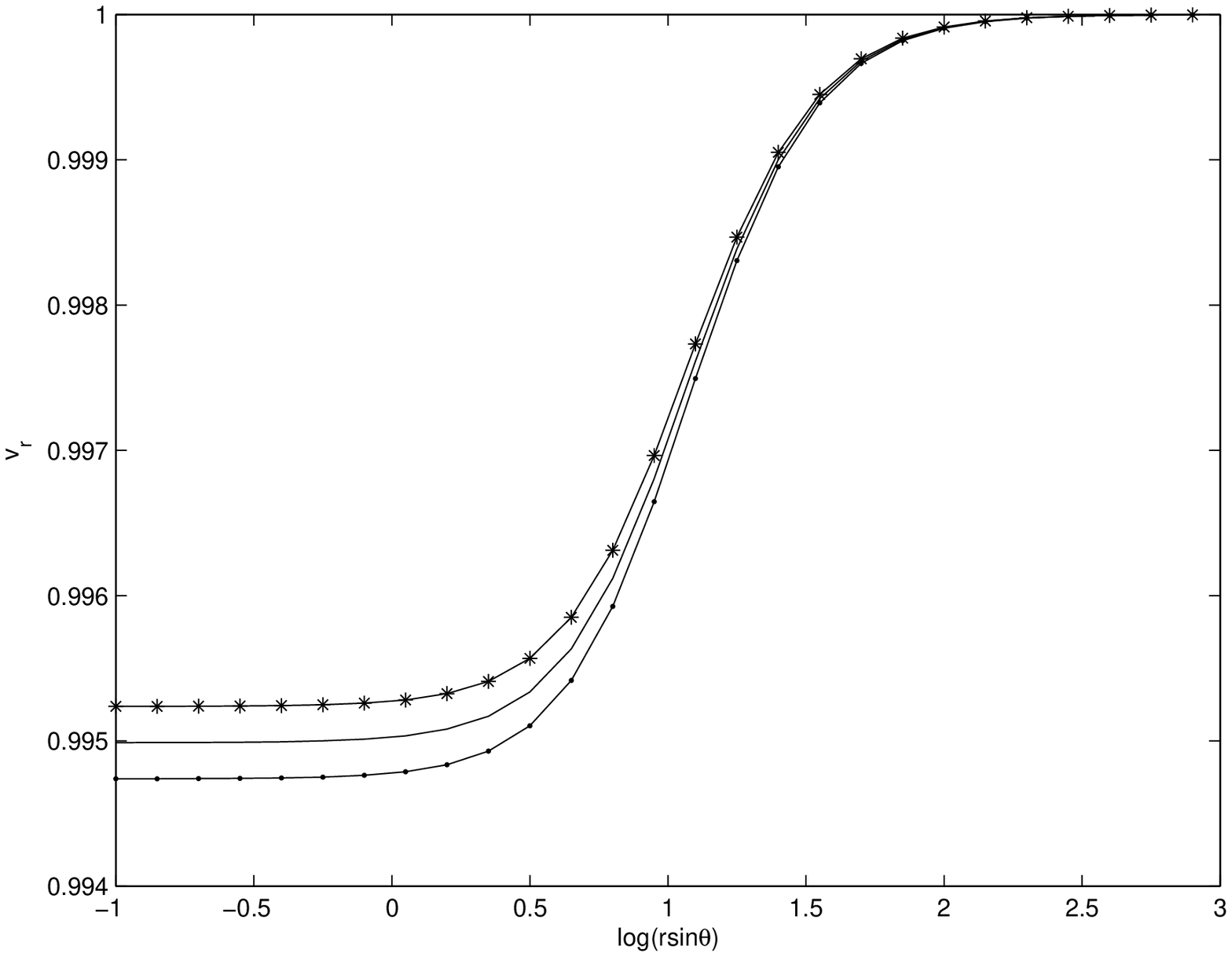}
\caption{Radial velocity component for electrons (asterisks) and positrons (points) as a function of the axial distance; $\varepsilon=0.1$, $\gamma_i=10$. The solid line without markers shows the quantity $w0$ determined by equation (\ref{eq25}).}
\label{f1}
\end{figure*}

Taking
\begin{equation}
\mu=1+\varepsilon,\quad \varepsilon\ll 1,
\label{eq29}
\end{equation}
one can present the particle velocities as
\[
v_{r_-}=w_0+\varepsilon w_1\,,
\]
\begin{equation}
v_{r_+}=w_0+\varepsilon w_1-\varepsilon w_0(1-w_0)\,,
\label{eq30}
\end{equation}
where $w_0$ is given by equation (\ref{eq25}), and, correspondingly,
\[
u_-=u_0+\varepsilon\frac{w_1}{(1-w_0)^2}\,,
\]
\[
u_+=(1-\varepsilon)u_0+\frac{\varepsilon w_1}{(1-w_0)^2}\,,
\]
where $u_0\equiv w_0/(1-w_0)$. Note that, in accordance with equation (\ref{eq24}), the case $\varepsilon\ll 1$ implies small first-order longitudinal electric field, $a\sim\varepsilon\ll 1$. To the first order in $\varepsilon$, equation (\ref{eq28}) is reduced to
\begin{equation}
t\frac{\mathrm{d}y}{\mathrm{d}t}=\frac{-2y^2+3y-1}{2y}+\frac{\varepsilon(1-3y/2)}{2y}\,,
\label{eq31}
\end{equation}
where $y\equiv u_+/t$. Solving equation (\ref{eq31}) in terms of series in $\varepsilon$ and taking into account the symmetry with respect to the simultaneous change $\varepsilon\to -\varepsilon$, $v_{r_\pm}\to v_{r_\mp}$, we come to the solution
\begin{equation}
u_\pm=u_0\left (1\mp\varepsilon/2\right )\,,
\label{eq32}
\end{equation}
where 
\[
u_0=\frac{t\sqrt{Ct+1}}{\sqrt{Ct+1}-1}
\]
and $C\equiv C_+=C_-=1/\gamma_i^2$, which translates into
\begin{equation}
v_{r_\pm}=w_0\mp\frac{\varepsilon}{2}w_0(1-w_0)\,.
\label{eq33}
\end{equation}
The velocities given by equation (\ref{eq33}) are plotted in Fig.~\ref{f1} and certainly coincide with the corresponding numerical solution of equation (\ref{eq28}). Along with the relations (\ref{eq12}), they present the self-consistent motions of the electron and positron constituents in the magnetosphere of a monopolar structure in case of vanishingly small inertia, $\xi\to 0$ and weak first-order longitudinal electric field, $a\sim\varepsilon\ll 1$.

\begin{figure*}
\includegraphics[width=190mm]{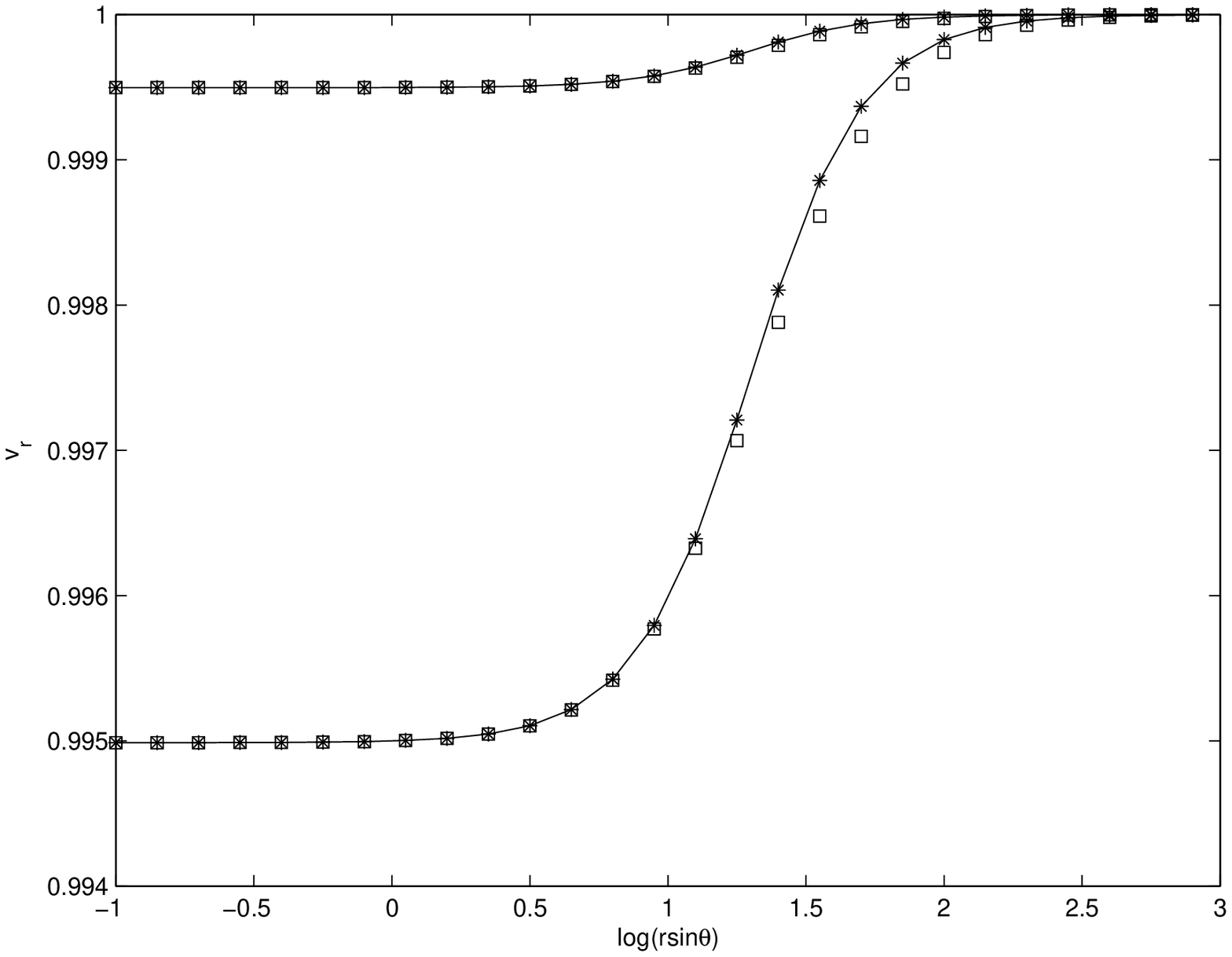}
\caption{The same as in Fig.~\ref{f1}, save that $\mu=10$. The squares show the approximate dependences corresponding to $u_\pm=u_0\mu^{\mp 1/2}$.}
\label{f2}
\end{figure*}

With equation (\ref{eq33}), the number densities of the particle species given by equation (\ref{eq15}) take the form $n_\pm\approx\rho/\varepsilon$, so that $\varepsilon^{-1}$ can be regarded as the plasma multiplicity, $\kappa\approx\varepsilon^{-1}$. As the standard models of pair creation in pulsars \citep[e.g.,][]{ha01,ae02} yield typical multiplicities $\kappa\sim 10-10^2$, the above assumption, $\varepsilon\ll 1$, is quite appropriate. At the same time, the values $\kappa\sim 1$ may also be the case. Note that, according to equation (\ref{eq33}), the relative difference of the particle velocities is $(v_{r_-}-v_{r_+})/v_{r_-}\sim(\kappa\gamma_c^2)^{-1}$, and it is believed to be always much less than unity, whereas the relative difference of the Lorentz-factors, $(\gamma_--\gamma_+)/\gamma_-\sim\kappa^{-1}$, may be of order unity. This corresponds to $(u_--u_+)/u_-\sim 1$ and $\mu\ga 1$. The numerical solution of equation (\ref{eq28}) for $\mu=10$ is presented in Fig.~\ref{f2}. One can see that the general character of the curves is the same as in Fig.~\ref{f1}. The plot also shows the dependencies $u_\pm=u_0\mu^{\mp1/2}$, which seem to approximate the numerical result with more or less admissible accuracy and can be used in further applications.

\begin{figure*}
\includegraphics[width=190mm]{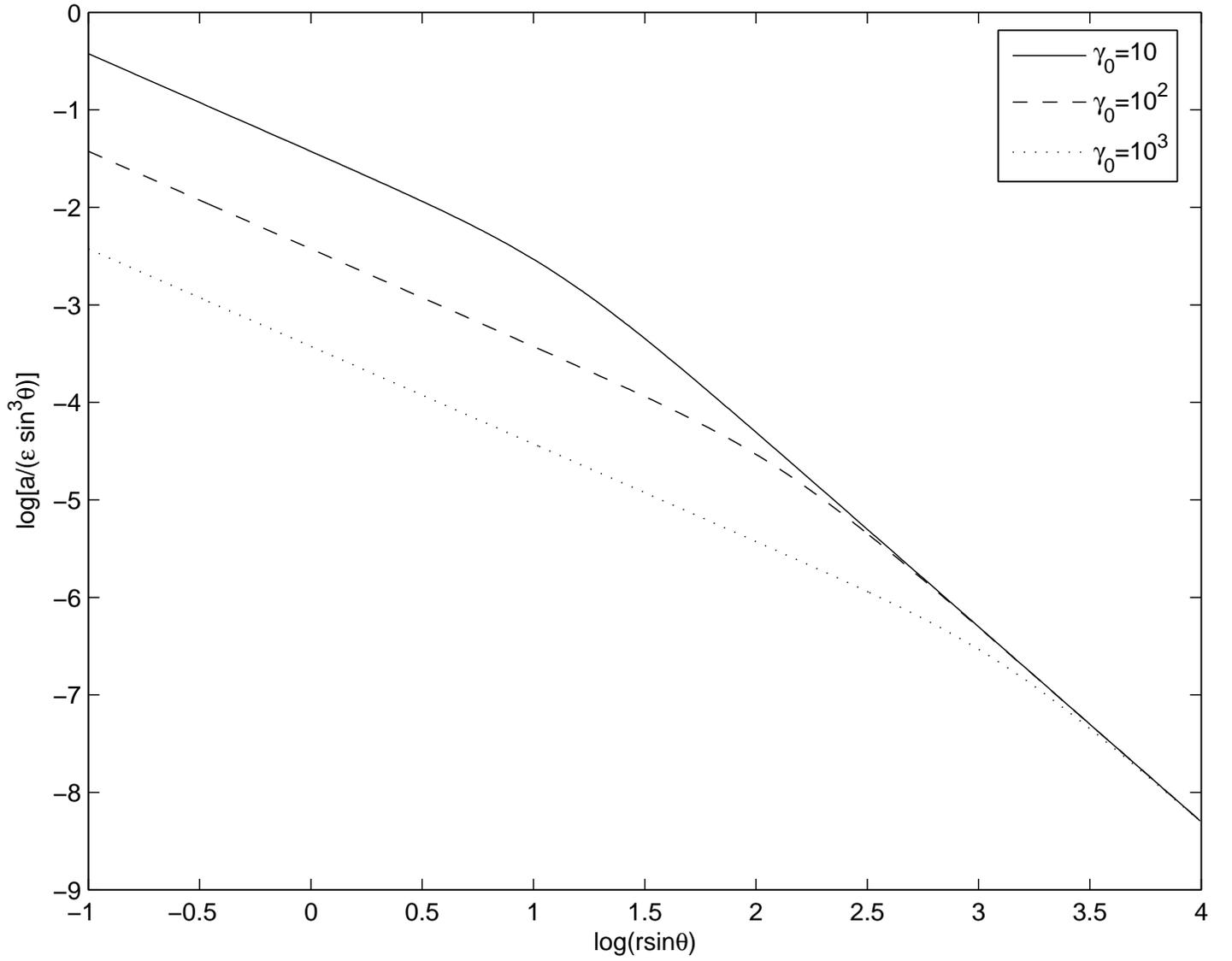}
\caption{Self-consistent first-order longitudinal electric field as a function of the axial distance for different initial Lorentz-factors.}
\label{f3}
\end{figure*}

And finally, one can derive the corresponding first-order longitudinal electric field. Rewriting equation (\ref{eq24}) in terms of $u_\pm$ leads to
\begin{equation}
u_\pm\frac{\mathrm{d}u_\pm}{\mathrm{d}x}-\frac{1}{2}(3u_\pm-x+1)=\pm(1+2u_\pm-x)^{3/2}\frac{a\sqrt{x}}{2\sin^3\theta}\,.
\label{eq34}
\end{equation}
Then, substituting equation (\ref{eq32}) into equation (\ref{eq34}), to the first order in $\varepsilon$ one obtains
\begin{equation}
a=-\varepsilon\frac{\sin^3\theta\sqrt{C}}{2\sqrt{t+1}}\frac{\sqrt{Ct+1}+2}{(1+\sqrt{Ct+1})^2}\,.
\label{eq35}
\end{equation}
Equation (\ref{eq35}) yields the inverse conductivity, $a\propto\sigma^{-1}$ as a function of the spatial coordinates $r$ and $\theta$, the particle initial Lorentz-factor, $\gamma_i=1/\sqrt{C}$, and the plasma multiplicity, $\varepsilon\approx\kappa^{-1}$. The dependencies for different initial Lorentz-factors are plotted in Fig.~\ref{f3}. The analogous curves for $\mu=10$ are presented in Fig.~\ref{f4}a, whereas Fig.~\ref{f4}b shows the curves for different $\mu$.

\subsection{Particle trajectories}

Now, with the zero-order velocities in hand, one can find the trajectories of the particles sustaining the force-free configuration. By definition,
\[
\frac{\mathrm{d}r_\pm}{\mathrm{d}\tau}=v_{r_\pm}\,,
\]
\begin{equation}
r_\pm\frac{\mathrm{d}\theta_\pm}{\mathrm{d}\tau}=v_{\theta_\pm}\,,
\label{eq36}
\end{equation}
\[
r_\pm\sin\theta_\pm\frac{\mathrm{d}\phi_\pm}{\mathrm{d}\tau}=v_{\phi_\pm}\,,
\]
where the velocities obey equation (\ref{eq12}). As $v_{\theta_\pm}\equiv 0$, the trajectories lie on the conical surfaces, $\theta=\mathrm{const}$. Excluding $\tau$ from the first and the third equations of the set (\ref{eq36}), one obtains
\begin{equation}
\mathrm{d}\phi_\pm=\frac{\mathrm{d}r_\pm}{u_\pm}\,.
\label{eq37}
\end{equation}
In case of $\varepsilon\ll 1$, $u_\pm$ is given by equation (\ref{eq32}) and integration of equation (\ref{eq37}) yields
\begin{equation}
\phi_\pm=\frac{1\pm\varepsilon/2}{\sin\theta}\mathrm{atanh}\frac{r\sin\theta}{\gamma_i\sqrt{r_\pm^2\sin^2\theta-1+\gamma_i^2}+\gamma_i^2-1}\,,
\label{eq38}
\end{equation}
where it is taken that $\phi_\pm=0$ at $r_\pm=0$. Keeping in mind equation (\ref{eq12}), from equation (\ref{eq36}) we have
\[
\frac{\mathrm{d}\phi_\pm}{\mathrm{d}\tau}=1-\frac{\mathrm{d}r_\pm}{\mathrm{d}\tau}\,,
\]
so that the law of particle motion reads
\begin{equation}
\phi_\pm+r_\pm=\tau\,,
\label{eq39}
\end{equation}
where it is taken that $r_\pm(\tau=0)=\phi_\pm(\tau=0)=0$.

Figure 5 shows the particle trajectories along with the magnetic field line at the same $\theta$. The cone generator at $\phi=0$ is added for the sake of demonstrating the distinction from the purely radial motion. The two close curves correspond to the case $\varepsilon\ll 1$ and are plotted based on equation (\ref{eq38}). The other two curves correspond to $\mu=10$ and are obtained using the numerical solution of equation (\ref{eq28}) in equation (\ref{eq37}). Note that the curvature of the trajectories is noticeable, though not too large.

\section{Discussion} 
\protect\label{s4}

Our consideration is concerned with the self-consistent two-fluid model of the pulsar magnetosphere. It is undoubtedly more informative than the customary force-free model. On the other hand, however, preceding numerical studies \citep{ko09} have demonstrated that in general it is too complicated to directly give transparent results. As is shown in Sect.~2.2, at the conditions relevant to the pulsar magnetosphere, the inertial effects are generally too weak to affect the basic quantities substantially. Therefore in the present paper we address the low-mass limit of the problem. Providing a reasonable simplification of the two-fluid model, our approach, at the same time, allows to reconstruct a realistic picture of the particle motions which sustain the force-free configuration. In case of negligible inertia, the customary radial speed-of-light motion appears only a particular case, whereas in general the particle velocities should obey equation (\ref{eq12}). The velocities themselves can be determined by taking into account vanishingly small inertia. They are obtained from the solvability condition of the first-order equation of motion. The resultant particle distributions enrich the customary force-free picture with new important features.

As is found in Sect.~3, it appears impossible to construct the self-consistent two-fluid model in the absence of the first-order longitudinal electric field. It is the quantity that determines the difference of the zeroth-order particle velocities. Hence, within the framework of our model, the force-free approximation bears imprints of the next-order particle acceleration in the longitudinal electric field. The corresponding finite parallel conductivity, $\sigma_\Vert\propto m^{-1}$, is expressed in terms of $v_{r_\pm}$ and seems to give a more realistic description of the acceleration processes in the pulsar magnetosphere as compared to the ad hoc conductivity of massless particles normally addressed in the literature. Note that our model can naturally incorporate not only the linear acceleration of the particle species. With the quantity $v_{1\theta_\pm}$ in hand, one would include the particle gyration as well. Although both the linear acceleration and particle gyration are as small as $\xi\gamma$ and their contribution to the general force-free picture is minimal, they still may be responsible for the observed high-energy radiation, whose level is typically much less than the total energetic output of the pulsar.

Thus, our model seems a proper context to examine any type of radiation processes by the accelerated particles and to interpret the pulsar high-energy emission. Furthermore, it also has important consequences for the pulsar radio emission. First of all, the relative motion of the particle species may cause the two-stream instability which is traditionally believed to underlie the pulsar radio emission mechanism. Of course, as is known since the paper of \citet{bb77}, the difference in the particle momenta should be at least as large as the temperature scatter. Although our consideration does not include the temperature effects and one cannot conclude as to the development of the two-stream instability in this case, it should be noted that actually our consideration does not place any restriction on the value of the shift in the distributions of the particle species. The shift is solely determined by the details of particle acceleration in the longitudinal electric field.

It is worth pointing out that the relation between the velocity shift and the accelerating electric field implies the physical connection between the resultant radio and high-energy emissions of pulsars. With the present state of art, the concrete observational consequences are still obscure, but it is important that this connection is an inseparable constituent of the global magnetospheric structure of the pulsar.

In our model, the particle motion is no longer radial. A slight curvature of the trajectories can be seen in Fig.~\ref{f5}. Although the corresponding azimuthal velocity components are small, the resultant differential rotation of the particles may have important implications. In particular, this hints at the possibility of the diochotron instability which is regarded as a mechanism of the radio subpulse phenomenon. A general idea and some illustrative calculations for ad hoc distributions were given in \citet{fkk06}. Although our present consideration does not provide sufficient details, one can expect that as both $\bmath{v}_\pm$ and $n_\pm$ depend on $\gamma_i(\theta)$, the proper distributions may really exist.

As is evident from Figs.~\ref{f1}, \ref{f2}, $v_{r_\pm}$ change with distance from the magnetic axis. Acceleration in the self-consistent fields occurs in such a way that at any point the electromagnetic force is compensated, the continuity condition is satisfied and the global charge and current are matched. At $r\sin\theta\sim\gamma_i$, a slight increase of the velocities is replaced by a drastic growth, in which case the low-mass approximation, $\xi\gamma\ll 1$, is ultimately broken. Apparently, it is the place where the energy of the self-consistent fields is intensely transmitted to the particle flow. Besides that, the closure of the global current circuit of a pulsar is also believed to occur there. Thus, our simplified model still predetermines the main problems of pulsar electrodynamics, such as the $\sigma$-problem and the current circuit configuration.

It should be noted that in our consideration the equation of particle motion (\ref{eq1}) does not include the radiation reaction force. In contrast to other neglected effects, such as gravitation or collisions, the radiation damping is believed to be very strong, especially close to the neutron star surface \citep[e.g.,][]{f89,gruz08}. For a probe particle moving in an external magnetic field, this would result in almost immediate loss of the perpendicular momentum, making the particle slide along the magnetic field line, in which case the radiation reaction force is zero. For an ultrarelativistic particle in arbitrary electric and magnetic fields, the radiation reaction force is reduced to \citep[e.g.,][]{f89}
\begin{equation}
\bmath{f}_{\mathrm{rad}_I}\propto\bmath{v}\left [\left (\bmath{E}+\bmath{v}\times\bmath{B}\right )^2-\left (\bmath{E}\cdot\bmath{v}\right )^2\right ]\,.
\label{eq40}
\end{equation}
Using equations (\ref{eq9})-(\ref{eq12}) in equation (\ref{eq40}), one can see that in case of negligible inertia $\bmath{f}_{\mathrm{rad}_I}$ is exactly zero. Furthermore, it does not contribute to the first-order equation of motion as well. Strictly speaking, one should also take into account another term of $\bmath{f}_{\mathrm{rad}}$, which is only $\gamma^2$ less than $\bmath{f}_{\mathrm{rad}_I}$. (Recall that in our consideration $\gamma$ is not a large parameter.) According to \citet{ll71},
\begin{equation}
\bmath{f}_{\mathrm{rad}_{II}}\propto\left [\left (\bmath{E}+\bmath{v}\times\bmath{B}\right )\times\bmath{B}+\bmath{E}\left (\bmath{E}\cdot\bmath{v}\right )\right ]\,.
\label{eq41}
\end{equation}
It is evident that this term also does not contribute to the zeroth-order equation of motion. And it is not difficult to see that incorporating equation (\ref{eq41}) into the first-order equation of motion by no means changes the solvability condition (\ref{eq24}), modifying only the definition of $a$. Thus, it is indeed reasonable to omit the radiation reaction force in our consideration. It should be kept in mind that the first- and higher-order components of the self-consistent electromagnetic fields are not obliged to supply particle motion along the trajectories dictated by the radiative damping, but the resultant gyration, being a higher-order effect, is beyond the framework of the present study.

It is important to note that the situation becomes substantially distinct as soon as the particle inertia is not negligible, i.e. at $r\sin\theta\sim\gamma_i$. Given that the inertial term is not small, the self-consistent electromagnetic fields change substantially, equation (\ref{eq11}) is no longer valid even approximately and $\bmath{f}_{\mathrm{rad}_{I,II}}$ are no longer cancelled. Thus, the inertia of strongly accelerated particles causes such self-consistent electromagnetic fields that imply strong radiation losses. The particle trajectories are believed to be markedly modified, and the particle gyration is no longer a small effect. All this may contribute to both the current closure and energy transfer from the fields to the particles.

In summary, taking into account vanishingly small inertia of the particles has a number of interesting consequences. Of course, the important restriction of our treatment is the assumption of a monopolar magnetic field, which is necessary for the transparent analytic treatment. However, our simplistic picture of the global magnetosphere of a pulsar still appears much more rich in physics than its preceding force-free analogues.

\section{Conclusions} 
\protect\label{s5}

We have considered the low-mass limit of the self-consistent two-fluid model for the magnetosphere of a monopolar structure. The solution obtained differs substantially from the well-known force-free picture, where the particles move radially at a speed of light. Now the number densities of the two particle species are definite, the velocities are distinct, both being less than the speed of light, and the particle trajectories demonstrate a slight curvature. The shift in the particle velocities is determined by the first-order longitudinal electric field, which appears the necessary ingredient of the self-consistent two-fluid model. Hence, the parallel conductivity of the plasma is $\propto m^{-1}$, as opposed to the customary conductivity of massless particles considered in the literature.

Our model incorporates the parallel acceleration of the particles, whose kinetic energy density is much less than the energy density of the magnetic field. Besides that, it allows to include particle gyration. Therefore it is believed to be a proper context to include the radiative processes into the global magnetosphere and to interpret the pulsar high-energy emission.

Our results may also have weighty implications for the pulsar radio emission. Difference in the velocities of the particle species may result in the two-stream instability which may give rise to the plasma waves convertible to the observed radio emission. The relation between the velocity shift and the longitudinal electric field is expected to underlie the connection between the radio and high-energy emissions of pulsars. Within the framework of our model, the particles perform differential rotation, which may cause the diochotron instability and be responsible for the radio subpulse phenomenon in pulsars.

Apart from acceleration in the first-order longitudinal electric field, the particle velocities change due to the spatial dependence of the zeroth-order self-consistent electromagnetic fields. As a result, $\bmath{v}_{0_\pm}$ increase with distance to the magnetic axis. At $r\sin\theta\sim\gamma_i$, a slight growth switches to a drastic one, so that the particle Lorentz-factors increase intensely and the low-mass limit, $\xi\gamma\ll 1$, is ultimately broken. Apparently, the locus $r\sin\theta\sim\gamma_i$ is the place where most part of the magnetic field energy is transmitted to the particles. One can expect that the closure of the global current circuit of a pulsar also occurs there. Of course, a stringent treatment of these issues, which are the basic problems of pulsar electrodynamics, is beyond the scope of our simplistic model.

On the whole, the self-consistent two-fluid model seems a promising description of the pulsar magnetosphere. The above considered low-mass limit for the monopolar magnetic field can be regarded as a starting point for analytic and numerical studies of the realistic magnetosphere, including not only its global structure, energy transport and current circuit configuration, but also the radiative processes, pair creation, etc.

\section*{Acknowledgements}
I am grateful to the anonymous referee for the suggestive questions.

\clearpage

\clearpage

\clearpage

\clearpage

\begin{figure*}
\includegraphics[width=190mm]{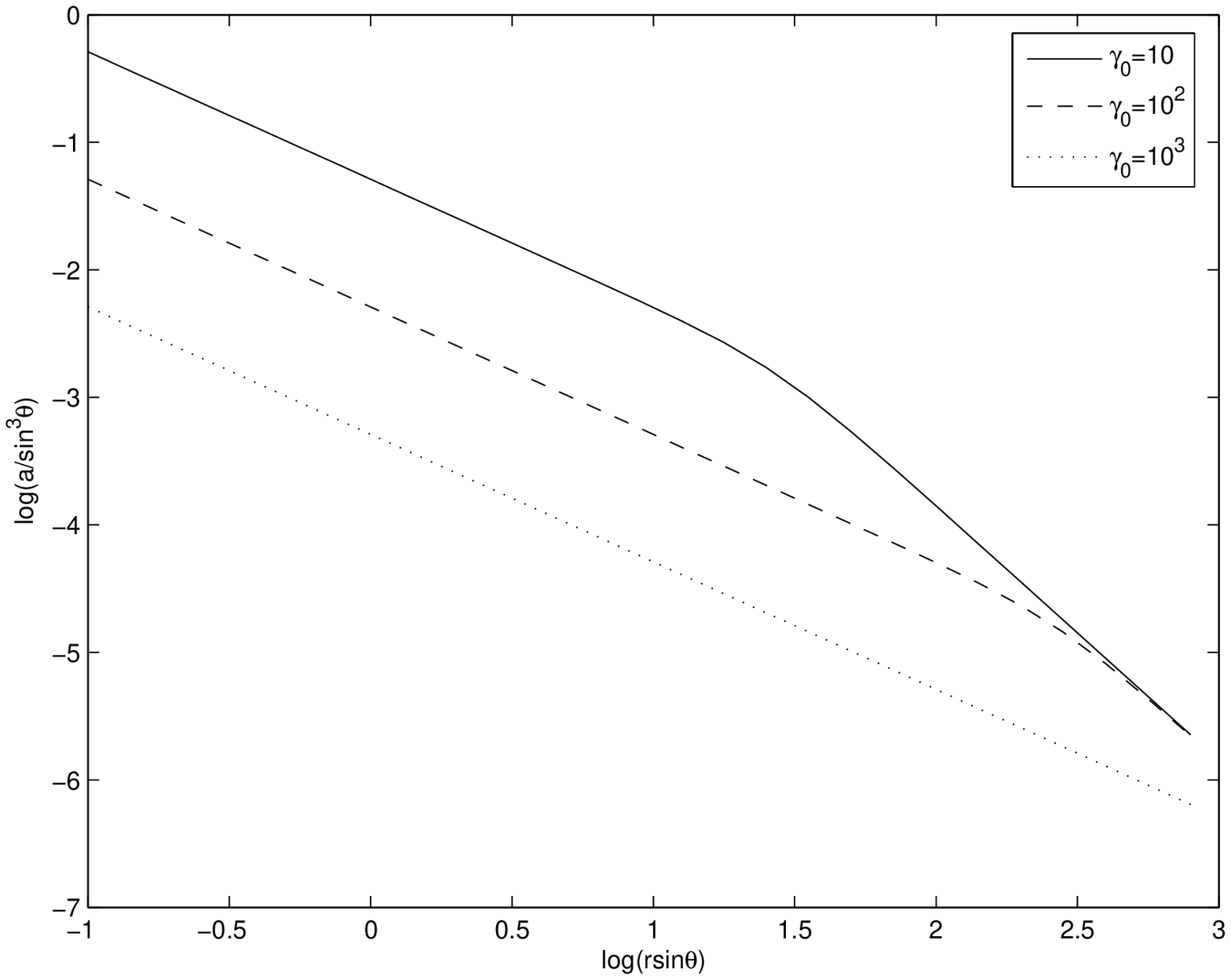}
\caption{The same as in Fig.~\ref{f1}, save that $\mu=10$. The squares show the approximate dependences corresponding to $u_\pm=u_0\mu^{\mp 1/2}$.}
\end{figure*}

\clearpage

\begin{figure*}
\includegraphics[width=190mm]{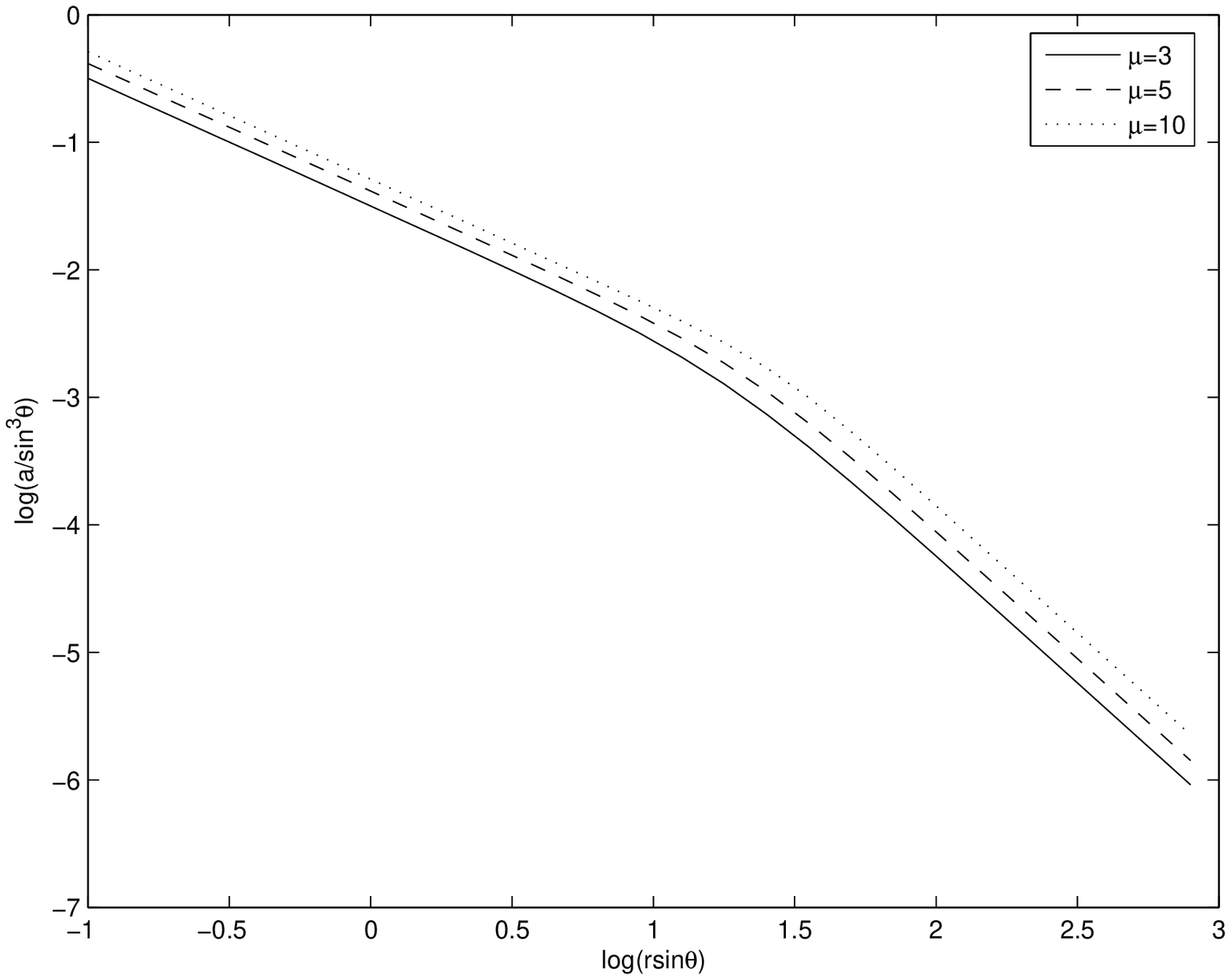}
\label{f4}
\end{figure*}

\clearpage

\begin{figure*}
\includegraphics[width=190mm]{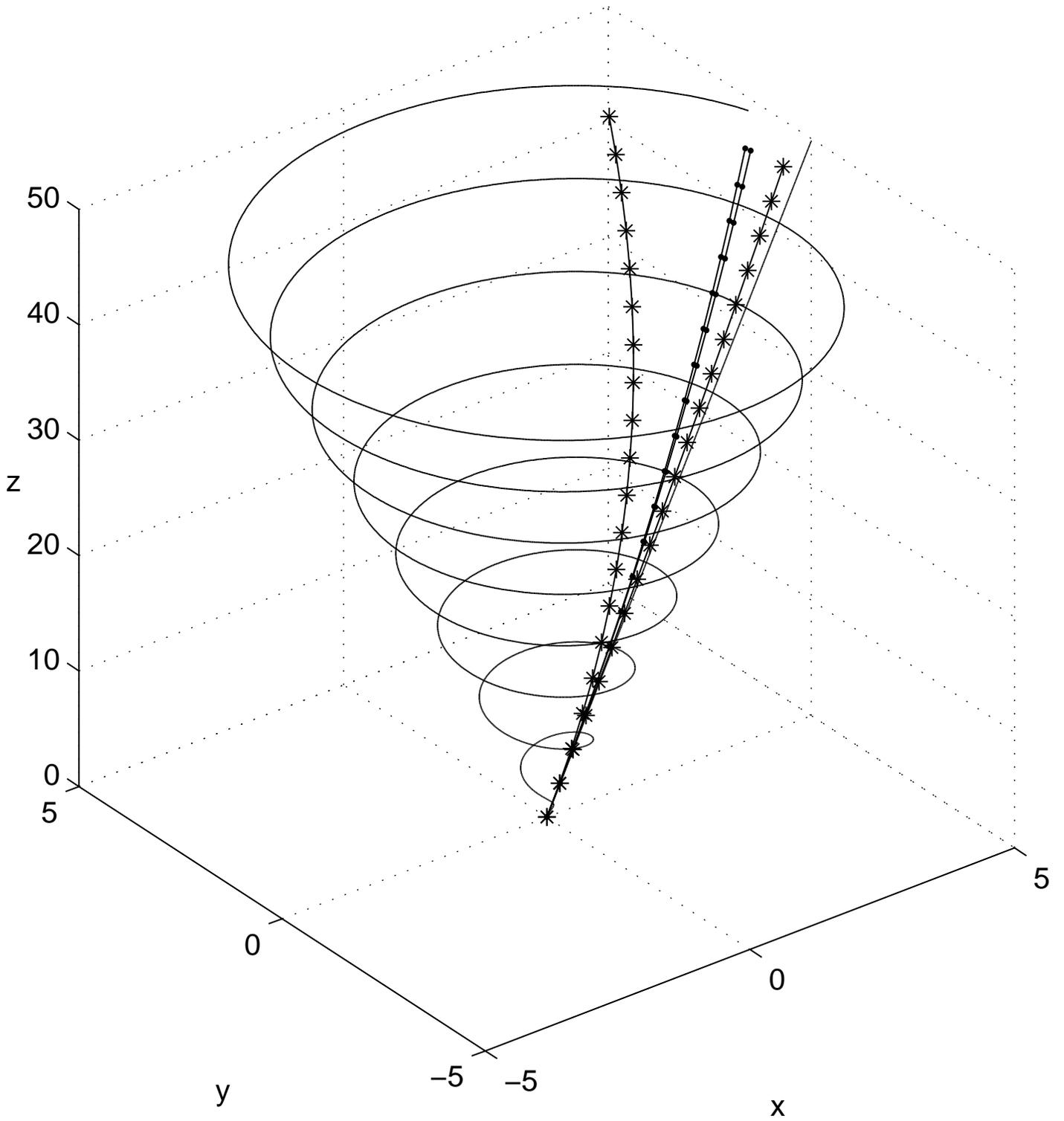}
\caption{Particle trajectories in the self-consistent flow; $\gamma_i=10$, $\theta=0.1$. The points correspond to $\varepsilon=0.1$, the asterisks to $\mu=10$. The spiral shows the magnetic field line and the straight line the generator of the conical surface at $\theta=0.1$, $\phi=0$.}
\label{f5}
\protect\label{lastpage}
\end{figure*}

\end{document}